\documentclass[a4paper,11pt,fleqn]{article}
\usepackage{latexsym}
\usepackage{graphicx}

\author{Sudhir R. Jain$^{(1)}$, Beno\^{\i}t Gr\'{e}maud$^{(2)}$, Avinash Khare$^{(3)}$ \\
$^{(1)}$ Theoretical Physics Division\\
Bhabha Atomic Research Centre \\
Mumbai 400094, India \\ \\
$^{(2)}$ Laboratoire Kastler Brossel\\
Universit\'{e} Pierre et Marie Curie \\
T12 E1, 4 place Jussieu \\
75252 Paris Cedex 05, France \\ \\
$^{(3)}$ Institute of Physics, Sachivalaya Marg\\
Bhubaneswar 751005, India}

\title{\bf Quantum Expression of Classical Chaos}

\date{}
\begin{document}
\maketitle

\begin{abstract}

Even as we understand for long that the world is quantal and buried
in it is  classical dynamics which is chaotic, finding eigenfunctions analytically  
from the  the Schr\"{o}dinger equation has turned out to
be a near-impossibility. Here, we discover a 
class of chaotic quantum systems  for
which we obtain some analytically exact eigenfunctions in  closed
form. This paves way to an exact 
classical and quantum mechanical treatment of 
chaotic systems. Furthermore, we bring out  connections, underlying the discovery,  between 
different areas of physics and mathematics related to universality 
in fluctuations observed in a wide variety of complex quantum systems.

\end{abstract}

\section{Introduction}

World around us is rather complex where nonlinear phenomena abound.
Nonlinearities give birth to chaos and make it impossible to predict
long-time dynamics {\it (1,2)}. Chaotic behaviour is characterized by the existence
of positive Lyapunov exponents, which determine the rate of exponential
separation of very close trajectories in the phase space of the system.
Upon casting the chaotic systems in a quantum mechanical framework,
impressions of chaos are found in variety of statistical properties
of energy levels and eigenfunctions {\it (3-5)}. 
The fluctuation properties of energy level
sequences of chaotic quantum systems agree very well with the results 
 in random matrix theory (RMT) {\it (6)}. These extensive studies have given place
to quantum chaology or what we call `statistical quantum chaos'.

Eigenfunctions and eigenvalues contain all the 
information about a time-independent quantum
system. Chaotic eigenfunctions have been studied in detail and there are
two main ideas around which the general understanding has evolved. According
to Berry {\it (7)}, for chaotic systems, eigenfunctions corresponding to  excited
states are conjectured to be well-represented as a random superposition
of plane waves. This conjecture has a lot of  numerical support 
on one hand,  and, is connected to statistical mechanics on the 
other {\it (8,9)}.
The other idea ensues from Heller's discovery {\it (10,11)}
of scarring
of eigenfunctions by
periodic orbits, where the probability density is considerably enhanced
on the periodic orbit in configuration or phase space. This discovery has 
helped in understanding how classical  periodic orbits form the underlying 
fabric for quantum states, and it helps in appreciating the beautiful nodal 
patterns and contour plots of chaotic eigenfunctions. Nevertheless, an 
analytical expression isn't obtained, and, Berry's conjecture does not 
help in seeing how coefficients in the superposition arrange to give 
patterns.   

The objective we set out for ourselves is to get analytically exact expressions
for chaotic eigenfunctions in closed form.  As we shall see, the 
resulting aesthetic beauty is due to  some `magic' connections {\it (12)} 
involving RMT and its relation to other topics in physics  and mathematics, including 
complex quantum  systems,  Riemann zeta function,  
exactly solvable many-body problems, 
partial differential equations and the Riemann-Hilbert problem {\it (13)}. 
These different 
areas are related to each other by the statistical properties of sequences characterizing 
them, like energy levels of nuclei and chaotic quantum systems, zeros of the Riemann 
zeta function, and eigenvalues of transfer matrices of disordered conductors.   
Moreover, joint probability distribution function (JPDF) of eigenvalues  
of random matrices is given by a statistical mechanics 
problem involving particles with Coulomb-like interactions {\it (14)}. 
The JPDF gives the probability with which eigenvalues $\{E_i\}$ are found 
in intervals $[E_i,E_i+dE_i]$; all the correlation functions giving various 
physical quantities follow from this. 
These problems belong to the  
general  class of exactly solvable statistical mechanics models. In turn, these models 
can be converted to problems in Hamiltonian dynamics where one can study the classical 
and quantal aspects. 

In this paper, we extend these connections to another class of random matrix
models
developed to explain intermediate statistics found in pseudointegrable systems, the 
Anderson model in three dimensions at the metal-insulator transition point and in 
certain problems  in atomic physics {\it (15-18)}. 
The sequences involved in these systems 
exhibit a behaviour intermediate to regular and chaotic.  
Once again, the eigenvalue distribution of such random matrices is related to an exactly 
solvable statistical mechanics model where the interaction is a screened Coulomb-like 
along with an additional three-body term {\it (19)}. In this article, the abovementioned  
model is mapped  into a 
problem in Hamiltonian dynamics in $d$ dimensions which is shown to 
display  chaos. Thus, the exactly solved many-body problem  
is non-integrable in the sense that there are lesser number of  
constants of the motion than the degrees of 
freedom. Quantum mechanically, some chaotic eigenfunctions are then found 
analytically even though the underlying classical dynamics at those energies is 
chaotic. This is illustrated through  Sections 2 to 4, 
the ensuing general perspective  is given in  Section 5. 

It is worth noting that the existence of Bose-Einstein condensation in a related  
one-dimensional many-body problem at zero temperature is 
proved recently {\it (20)}. To prove the existence of a Bose-Einstein condensate in one dimension  at non-zero 
temperatures, we need the excited states. 
Thus, the development presented here is of importance to the general theory of quantum phase transitions.         
 
For  quantum cat maps (toral automorphisms),
analytical form for the eigenfunctions is known {\it (21,22)},
where the  solutions were possible because the semiclassical
studies turned out to be exact.
These are very interesting results
but the methods are so specialized that they do not throw  much light
on other related problems. In Section 5, we discuss the possible advances 
emerging from the discussions that follow in the context of general 
relevance of the results and the theme of this article.  

\section{From Many-Body Problem to One-Body Problem}

In 1999, a many-body problem in one dimension was discovered {\it (19,23)}
where the
nearest neighbours interact via a repulsive interaction which is inverse-
square in distance between the particles and an attractive three-body
interaction, also inverse-square in position coordinates of the
particles. The $N$-particle problem on a circle has the Hamiltonian :
\begin{eqnarray}
H &=& \sum_{i=1}^{N} \frac{p_i^2}{2m} +
g\frac{\pi ^2}{L^2} \sum_{i=1}^{N} \sin ^{-2} \left[
{\pi \over L} (x_i - x_{i+1})\right] \nonumber \\
&-& G \frac{\pi ^2}{L^2} \sum_{i=1}^{N}
\cot \left[{\pi \over L}(x_{i-1} - x_i) \right]
\cot \left[{\pi \over L}(x_{i} - x_{i+1}) \right]~,
\end{eqnarray}
with $x_j = x_{N+j}$.
In the rest of this article, we will take the mass as unity and the
circumference $L=\pi$, for notational simplicity.  The potential is singular
whenever $x_{j+1} = x_j + n\pi $ ($n$ being an integer)
with an inverse-square
singularity. This leads to disconnected domains where the wavefunctions
are zero at  singular (hyper-)planes in the quantum problem.
We choose the domain in which the
particles are  ordered as $x_1  \leq x_2 \leq ... \leq x_1 + \pi$.
The center of mass (CM) motion can be separated in this case by writing
the amplitudes of the motion of particles around the CM,
$X = {1 \over N}\sum_{i=1}^{N} x_i$ in terms of normal mode coordinates
{\it (24)}.
Thus, we can write the positions as $x_j = X + y_j$, and $y_j$'s as
\begin{eqnarray}
y_j &=& -{\pi \over 2} 
+ \left(j-{1 \over 2}\right){\pi \over N} + \frac{1}{\sqrt{N}} q_{M} \cos (\pi j)
\nonumber \\ 
&+& \sqrt{\frac{2}{N}} \sum_{n=1}^{M-1}
\left[q_n \cos \left(\frac{2 \pi n j}{N} \right) -
q_{-n} \sin \left(\frac{2 \pi n j}{N} \right) \right], 
\end{eqnarray}
for even $N$ ($M = N/2$); and
for odd $N$, without the last term on the right hand side and summation
going upto $n=M$.
The quantum Hamiltonian operator transforms to
\begin{eqnarray}
H &=& H_{CM} (X) + H_{\mbox{\small billiard}} (\{q_n\}) \nonumber \\
&=& -\frac{1}{2N}\frac{\partial ^2}{\partial X^2}
- \frac{1}{2}\sum_{n}^{M}\frac{\partial ^2}{\partial q_n^2} +
\sum_{j=1}^{N} W[y_j(\{q_n\})],
\end{eqnarray}
where $W[y_j(\{q_n\})]$ is the potential term in (1).
Note that
$n=\pm 1, \pm 2,...,\pm M$ for odd $N$ while $n = \pm 1, \pm 2,...,
\pm (M-1), M$ for even N. Thus,
$H_{\mbox{\small billiard}}$ represents a single particle in an ($N-1$)-
dimensional domain bounded by the (hyper-)planes where potential becomes
singular
forcing all the wave functions to be zero on the boundaries
of the domain. This gives us a
class of billiards as a function of $\beta $ (defined through $G = \beta ^2$
and $g = \beta (\beta - 1)$)
and from dimensions two to
$(N-1)$. A similar connection was realized by Rey and Choquard
\cite{rc} when they
considered the Calogero-Sutherland-Moser (CSM) system and showed that the
$N$-body problem in this case is mapped to an integrable billiard problem
in $N-1$ dimensions. It is interesting to note
that the CSM and our model coincide for three particles as nearest neighbours
are all the neighbours and the cotangent term is unity owing to a trigonometric
identity.
Unlike this, and most interestingly, our many-body
problem leads to non-integrable billiards for particles greater than
three and hence, billiards of dimension greater than two. We now turn to
show that for $N >3$, our billiard models are, in fact, chaotic.
This also implies non-integrability of the many-body problem.

\section{Family of Classical Billiards}

For the
sake of concreteness, we concentrate our discussion on
$N = 4$, leading to a three-dimensional
billiard. Equations of motion are singular each time a collision
occurs (involving 2, 3 or 4 particles), but two-particle collisions
are forbidden by energy
conservation, giving rise to smooth integration using standard
Runge-Kutta method, until one reaches a multiple collision
{\it (25)}. Preliminary studies show that regularization of the
classical motion near these collisions could be done, but its
exact implementation has not been achieved yet. Nevertheless, we
believe that this problem does not alter results presented in this section.
For each
trajectory, we have also computed the associated monodromy matrix, $M$
(whose symplectic structure is used as a relevant test), allowing us to
extract evidence of chaotic behaviour in our system. For this, we have
computed the ``Liapunov exponents'',
$\tilde{\lambda}(T)=\ln{(|M(T)\cdot  \mathbf{e}_0|)}$ using three different
vectors~: $\mathbf{e}_0^{\parallel}$, $\mathbf{e}_0^{\perp}$
and $\mathbf{e}_0^{\mathrm{r}}$, which are, respectively,  unit
vector parallel to the flow at an initial time,  unit vector
perpendicular to the energy shell at an initial time and a ``random''
unit vector, namely $1/\sqrt{6}(1,1,1,1,1,1)$.
$\tilde{\lambda}^{\parallel}(T)$, being equal to
$\ln{|\dot{\mathbf{X}}(T)|/|\dot{\mathbf{X}}(0)|}$, is
bounded entailing thereby a vanishing Liapunov exponent, providing
a reference scale for further numerical estimation of non-vanishing
Liapunov exponent. For $\beta=2$,
at classical energy equal to  the quantum ground state energy 
$\epsilon_0=4\beta^2$
(see Eq.~(\ref{4})), and initial conditions,
\begin{equation}
\left(\begin{array}{r}
q_1(0) \\
q_{-1}(0) \\
q_2(0) \end{array}\right)
=\left(\begin{array}{r}
0.1000 \\
0.2000 \\
0.0000\end{array}\right)
\quad
\left(\begin{array}{r}
p_1(0) \\
p_{-1}(0) \\
p_2(0) \end{array}\right)
=\left(\begin{array}{r}
1.0000 \\
2.0000 \\
5.1844 \end{array}\right)
\end{equation}
results on Liapunov exponents are plotted in Fig.~\ref{liapunov}. As
expected, behavior of
$\tilde{\lambda}^{\parallel}(T)$ is substantially different from those of
$\tilde{\lambda}^{\perp}(T)$  and $\tilde{\lambda}^{\mathrm{r}}(T)$,
emphasizing thus
the presence of hard chaos in the system, with Liapunov exponent,
$\lambda\sim 6$ {\it (26)}.


A more appealing evidence of hard chaos is obtained on plotting
Poincar\'e surface of sections (PSOS). For a generic 3D time-independent
system, their dimensionality (4D) makes them quite useless for visualising.
Fortunately, in our case, symmetry properties of the
Hamiltonian, viz., invariance
under $q_1\leftrightarrow q_{-1}$ exchange allows us to consider the
reduced phase space made of trajectories for which $q_1=q_{-1}$ and
$p_1=p_{-1}$ at any time, leading thus to an effective 2D
system. Fig.~\ref{PSOS} depicts, for $\beta=2$, the reduced PSOS
defined by $q_2=0$, at different energies~: $E_a=\epsilon_0=16$ (ground state
energy, see Eq.~(\ref{4})),
$E_b=\epsilon_1=25.5$ (see Eq.~(\ref{4.5})),
$E_c=\epsilon_N=36$ (see Eq.~(\ref{4.6})) and $E_d=100$.
The empty areas
appearing on all these plots correspond to trajectories having a four
body collision in their past, this has been checked by propagating
the corresponding initial conditions backward in time {\it (27)}.
Nevertheless the
system appears to be fully chaotic in the reduced phase
space, which is emphasized by the fact that all periodic orbits of the
reduced dynamics are unstable (i.e., non-trivial eigenvalues of the
monodromy matrix are not on the unit circle). Actually these orbits
are also unstable
when considered in the full phase space. Moreover, there are also
unstable periodic orbits not belonging to the reduced phase space.
Even more, we find unstable periodic orbits for
which the four non-trivial eigenvalues of the monodromy matrix form a
quadruplet $(\Lambda,\Lambda^*,\Lambda^{-1},\Lambda^{-1*})$, $\Lambda$
being complex. For example, for $\beta=2$, there is a periodic orbit of
length $1.1083386$ with initial conditions
(classical energy is $\epsilon_0$)~:
\begin{equation}
\left(\begin{array}{r}
q_1(0) \\
q_{-1}(0) \\
q_2(0)
\end{array}\right)
=\left(\begin{array}{r}
 -0.228018 \\
 -0.261062 \\
  0.000000
\end{array}\right)
\quad
\left(\begin{array}{r}
p_1(0) \\
p_{-1}(0) \\
p_2(0)
\end{array}\right)
=\left(\begin{array}{r}
  2.06029  \\
  0.737421 \\
  5.07910
\end{array}\right)
\end{equation}
for which we find $\Lambda=-367.64192 + \iota 64.633456$,
this is shown in Fig.~\ref{PO}.

$\beta=2$ is not a special value, we have checked that all results
presented in this section hold for other values $\beta>1$, with eventually
apparition of a mixed dynamics for large $\beta$.
A complete study of  classical dynamics has to be done, especially a
systematic search for periodic orbits and their properties (variations
with both $\beta$ and energy, bifurcations, and so on).


\section{Family of Quantum Billiards}

We shall now use the known exact energy eigenstates {\it (28)}
of the newly
discovered $N$-body Hamiltonian (1) and obtain few energy eigenvalues and
eigenfunctions of the corresponding $(N-1)$-dimensional billiard.
The billiard eigenfunctions are obtained by
eliminating the center-of-mass dependence while the eigenvalues are determined
by subtracting off the center-of-mass energy.

Let $E_k, \psi_k$ denote the energy eigenvalues and eigenfunctions of the
$N$-body problem (1) with periodic boundary conditions, i.e.
$H\psi_k = E_k \psi_k$. Then the exact ground state is given by {\it (19,23)}
\begin{equation}\label{4}
\psi_0 = \prod_j^N \mid \sin (x_j - x_{j+1}) \mid^{\beta}~,~~ E_0 = N\beta^2~,
\end{equation}
provided  $g, G$ are related by $g=\beta(\beta-1),
G=\beta^2$.

As mentioned in the Introduction,  $\psi _0^2$ exactly gives the analytical form
of the JPDF of eigenvalues of
random matrix ensemble relevant for intermediate
statistics {\it (15,18,19)} observed in plane polygonal billiards
{\it (15,29)}, Aharonov-Bohm billiards {\it (17)},
Anderson model in three dimensions {\it (16)}, and so on. 

In addition to the ground state, a few of the excited  energy eigenstates have also been
obtained in this case {\it (28)} and are given by ($\psi_k = \psi_0 \phi_k$)
\begin{eqnarray}\label{4.1}
\phi_1 &=& e_1~, ~~E_1 = E_0 + 2+4\beta~, \nonumber \\
\phi_{N-1} &=& e_{N-1}~, ~~E_{N-1} = E_0 + 2N-2+4\beta~, \nonumber \\
\phi_N &=& e_1 e_N -\frac{N}{1+2\beta} e_N~,~~E_N = E_0 + 2N+4+8\beta~.
\end{eqnarray}
Here $e_j$, $(j=1,2,...,N)$ is an elementary symmetric function of order $j$ in
the variable $z_j$. For example $e_2 = z_1z_2
+z_1z_3+...$ (containing $N(N-1)/2$ terms) and $z_j = \exp (2\iota x_j)$.
Note that $\phi_k$ is an eigenfunction of the momentum operator
with eigenvalue $k$ {\it (30)}. 
Following the treatment in {\it (24)} for the CSM
billiard, it is  easily shown that if the billiard Hamiltonian
satisfies the eigenvalue equation $H_B \chi_k = \epsilon_k \chi_k$, then
the eigenvalues $\epsilon_k$ and the eigenfunctions $\chi_k$ are related
to $E_k, \psi_k$ by
$\chi_k = \exp (-2\iota k X) \psi_k$ with $\epsilon_k = E_k -\frac{2k}{N}$.
Finally, we then obtain the following excited
eigenvalues and eigenfunctions for the $(N-1)$-billiard
\begin{equation}
\chi_{1c} =  \psi_0 \bigg [\cos \frac{2}{N}([N-1]x_1-x_2-...-x_N) +
\mbox{cyclic permutations} \bigg ],
\end{equation}
\begin{equation}
\chi_{1s} =  \psi_0 \bigg [\sin \frac{2}{N}([N-1]x_1-x_2-...-x_N) +
\mbox{cyclic permutations} \bigg ],
\end{equation}
\begin{equation}\label{4.5}
\epsilon_{1c} = \epsilon_{1s} = E_0 +4\beta +2 -\frac{2}{N}~,
\end{equation}
\begin{equation}
\chi_{N} =  \psi_0 \sum_{i<j=1}^{N} \cos 2(x_i -x_j) +\frac{N\beta}{1+2\beta}~,
\end{equation}
\begin{equation}\label{4.6}
\epsilon_N = E_0 +8\beta +4~,
\end{equation}
where $E_0$ is as given by Eq. (\ref{4}). Note that even though $\chi_N$ is
$N$-dependent, the energy difference $\epsilon_N - E_0$ is $N$-independent. 

These expressions conclude our illustration of the connections between different 
areas of research in physics and mathematics, as noted in the Introduction and 
by Kadanoff {\it (12)}. The classical dynamics at energy equal to the energy 
eigenvalues of the eigenfunctions is chaotic, yet the eigenfunctions are not
random
superposition of plane waves. However, since $\beta $ can assume any value, the 
functions could be quite complicated, yet comprehensible.

\section{Summary }

We underline a fundamental theme which is at the heart of
our findings.  Let us first present 
two scenarios - {\it Scenario 1:} From pseudointegrable
systems to a class of fully chaotic systems, and, {\it Scenario 2:} From
fully chaotic systems to a class of integrable systems.
\vskip 0.25 truecm
\noindent
{\it Scenario 1}\\

Let us begin with classically pseudo-integrable billiards which are
neither chaotic (zero Kolmogorov-Sinai (KS) entropy) nor integrable.
On quantization, we get the energy levels
and eigenfunctions. The local statistical properties of energy levels
agree well with the Short-Range Dyson Model (SRDM).
The JPDF of eigenvalues is such that its analytical
form coincides with the square of the ground-state wavefunction of the
many-body problem (Cf. Section 2). As shown in Section 2, this many-body
problem is mapped to a family of quantum billiards. The classical
analogues of these quantum billiards are fully chaotic.

\vskip 0.25 truecm
\noindent
{\it Scenario 2}\\

We begin with
classically chaotic systems and note that on quantization, the local
fluctuation properties agree well with those of canonical RMT. This
is very well-studied. The JPDF of eigenvalues of the random matrix
ensembles is such that its analytical form coincides with the square
of the ground-state wavefunction of the CSM.
In the CSM, there are $N$ particles on a circle
and each pair interacts through an inverse-square potential. This is a
completely integrable system. Since the ground-state is crystalline, here
again the mapping to billiards in $(N-1)$ dimensions is possible.
However, in this case, the billiards are fully integrable for any $N$.

\vskip 0.25 truecm

Thus, we see that the two random matrix models - canonical and the SRDM -
both have two arms, one arm leads to chaos ($h_{KS} > 0$) and the
other to order ($h_{KS} = 0$). The arm which leads to analytical
results for chaotic systems belongs to the SRDM and the associated
many-body problem due to Jain and Khare {\it (19)}. Thus, we  suggest that
for quantizing chaotic systems, the route through many-body problems
may be more worthwhile. Of course, it would require a great deal of
ingenuity to construct such many-body problems. However, the recent work
of Glashow and Mittag {\it (31)} shows that it is possible for any triangular
billiard, where the authors generalise an old connection of Onsager
and Sinai {\it (32)}.

There is another important future direction which suggests itself from
the Scenarios presented above. The quantum systems that we usually deal
(billiards, for instance) with, have an infinite-dimensional Hilbert space.
Thus the Hamiltonian matrix is infinite-dimensional. 
This means that the
random matrices also should be of an infinite order, which manifests as
universality in the scaling limit.
This would lead to
many-body problem with infinite degrees of freedom - known as  fields 
or classical fluids governed by partial differential equations.   
An instance of relevance
here is the Korteweg-de Vries equation associated with the CSM. It is
only the truncations and projections of these field solutions and the
vast variety of solutions admitted by the partial differential equations
that manifest themselves in the standard parlance and certainly in
numerical works. As discussed by Deift {\it (13)}, the solvability of 
certain partial differential equations demands the solution of the 
Riemann-Hilbert problem which, in turn, brings in matrix theory. We
believe that herein underlie  some deep secrets of physical
theories. 

Finally, a concluding remark on quantum theory of chaotic systems.
We have seen
above that low-lying eigenfunctions built on classical chaos - termed as
`Chaotic States' -  are not random
superpositions of
plane waves. Thus, Berry's conjecture would be expectedly true
for highly excited
chaotic states, and not   for chaotic states in general. 
We believe that chaos in quantum
wavefunction, even at low energies, would show up as the system evolves.
The time-dependent wavefunction, written as a superposition of
eigenfunctions, has coefficients displaying chaos. Thus, eigenfunctions
form an invariant
set in quantum theory,  much in the same way as periodic orbits
and fixed points do in 
classical theory. The most important evidence is shown by
Heller's discovery
of scars. Thus, we state - scarring of wavefunctions
on periodic orbits
is a reminder of the classical fact that on its way,
an arbitrary trajectory is shadowed by periodic orbits.

Authors thank D. R. Arora, D. Delande, J. Robert Dorfman, A. Mitra, 
S. Nonnenmacher, A. K. Pati, and J. Wm. Turner for scientific discussions 
and numerous suggestions which led to the present form of the article.    
The work done by AK forms a part of the Indo-French Collaboration Project
1501-1502. 

\newpage
{\large \bf Figure Captions}
\vskip0.5cm
\begin{itemize}
\item[1]{$\tilde{\lambda}^i(T)=\ln{(|M(T)\cdot
\mathbf{e}^i|)}$, for $i=\parallel$ (continuous line), $\perp$
(dotted line) and $\mathrm{r}$ (dashed line), for $\beta=2$ (see text for
definition and initial conditions). As expected,
$\tilde{\lambda}^{\parallel}(T)$ is bounded, whereas
$\tilde{\lambda}^{\perp}(T)$ and $\tilde{\lambda}^{\mathrm{r}}(T)$
show linear
behavior, emphasizing  presence of hard chaos in the system.}
\\
\item[2]{Reduced PSOS
defined by $q_2=0$, at different energies~: $E_a=\epsilon_0=16$ (ground state
energy, see Eqs.~(\ref{4})),
$E_b=\epsilon_1=25.5$ (see Eq.~(\ref{4.5})),
$E_c=\epsilon_N=36$ (see Eq.~(\ref{4.6})) and $E_d=100$. The reduced
dynamics is clearly chaotic. (See text for discussion about areas
appearing in the PSOS).}
\\
\item[3]{Positions (top plot) and momenta (bottom plot) as
function of time for
an unstable periodic orbit of the system with $\beta=2$ at classical
energy $E=\epsilon_0$ (ground state energy). Continuous line:
$(q_1,p_1)$, dotted line  $(q_{-1},p_{-1})$ and dashed line
$(q_2,p_2)$. The four non-trivial eigenvalues of the monodromy
matrix forms the quadruplet
$(\Lambda,\Lambda^*,\Lambda^{-1},\Lambda^{-1*})$, with
$\Lambda = -367.64192 + \iota  64.633456$}
\end{itemize}


\begin{thebibliography}{99}
\bibitem{note1} Occurrences of the latter case are systematically
  checked in our code and the corresponding trajectories are
  discarded.
\bibitem{numnote} From a numerical point of view, one should mention that
the present trajectory alone can be followed up to time $T\sim80$
without trouble and also that at $T=4$, the symplectic structure has
been preserved better than $10^{-8}$ (relative). Of course, these
behaviors are not qualitatively modified on
changing either initial conditions or energy.
\bibitem{foot} More precisely,
trajectories having initial conditions in the two largest areas  will
perform the collision (backward in time) before returning to PSOS,
whereas other areas are images of these largest areas by the Poincar\'e
map (backward in time).
\end{thebibliography}
\end{document}